\documentclass[prl,twocolumn]{revtex4}
\usepackage{amsmath}
\usepackage{graphicx}

\begin{document}

\title{Diverging Static Length Scales and Clustering in Cold Lennard-Jones Fluids }

\author{Paul C. Whitford}

\affiliation{Department of Physics, University of California, San Diego, La Jolla, CA 92093}

\author{George D. J. Phillies}
\email[To whom inquiries should be sent ] {phillies@wpi.edu}

\affiliation{Department of Physics, Worcester Polytechnic
Institute,Worcester, MA 01609}

\begin{abstract}

We report molecular-dynamics simulations on a three-dimensional two-component Lennard-Jones fluid.  We identify two distinct static length scales.  The longer length scale, which diverges at low temperatures, is continuous with the high-temperature static correlation length. The shorter length scale reflects static clusters that form below a $T_{mc} \approx 2$ well above an apparent melting point $T_{m} \approx 1.08$. We partially visualize the clusters;   their maximum radius is temperature-independent.  Cluster static and dynamic properties match Kivelson's model (S. A. Kivelson, et al., J. Chem.\ Phys.\ {\bf 101}, 2391 (1994)) for frustration-limited cluster-forming glasses.

\end{abstract}

\maketitle

Many liquids, when cooled below their melting points, transform not into crystals but into amorphous glasses.  There remains disagreement: Is glass formation an equilibrium thermodynamic transition or a dynamic effect? At a  thermodynamic transition, static atomic position correlations would change markedly.  Searches\cite{ernst,leheny} of the radial distribution function $g(r)$ of cold glass-forming liquids found no such changes.  More recent studies\cite{glotzer1,glotzer2} of multipoint \emph{dynamic} correlation functions find increasing position correlations between particularly mobile or immobile atoms as the glass is approached. 

We performed molecular dynamics simulations on a three-dimensional Lennard-Jones fluid.  In addition to the well-known local ($r \leq 5$ atomic diameters) order, \emph{we show that $g(r)$ has extended (out to $r > 11$) order in cold systems.}  The crests and troughs of $g(r)$ share a common envelope function; the function's length scale \emph{diverges} as $T \rightarrow 0$.  On extracting the extended order, cold Lennard-Jones fluids are found to form locally-ordered structures ("clusters"), whose density is that of the bulk, and whose maximum radius is temperature-independent.

Our fluid systems contained $N =125,000$ atoms, equally of species A and B, in an $L \times L \times L$ ($L \approx 43$) box with periodic boundary conditions. The number density was $\rho = N L^{-3} = 1.30$. The potential energy was
\begin{equation}
   U_{ij}  = 4 \epsilon [(\frac{\sigma_{ij}}{r})^{12} -(\frac{\sigma_{ij}}{r})^{6}],
    \label{eq:Uij}
\end{equation}
with $r$ the interparticle distance, $i$ and $j$ labelling species, and $U_{ij} = 0$ for $r>2.5$.  Following Glotzer, et al.\cite{glotzer}, we took $\sigma_{AA} = 1$, $\sigma_{BB} = 5/6$, and  $\sigma_{AB} = 11/12$.  The interparticle energy $\epsilon$ and particle mass $m$ were the same for all atoms.    In natural units the temperature $T$ is $k_{B}/\epsilon$ and the time unit is $\sigma_{AA} (m/\epsilon)^{1/2}$.   Numerical integration of Newton's equations of motion was accomplished via symplectic integration using the Calvo and Sanz-Serna fourth order method\cite{calvo,gray}. The time step was never greater than 0.01, and less at $T > 1$.  Simulations were made seriatim at 34 temperatures by progressive cooling and re-equilibration starting from an equilibrated $T=47800$ fluid. 

We calculated specific radial distribution functions $g_{AA}(r)$, $g_{AB}(r)$, $g_{BB}(r)$, and the total radial distribution function $g(r)$, with normalization $g_{ij}(\infty) = 1$, which give likelihoods of finding two atoms separated by $r$.  Following Steinhardt, et al.\cite{steinhardt}, we also calculated spherical harmonic components
\begin{equation}
    Q_{LM}  = n^{-1} \sum_{i=1}^{n} Y_{LM} (\Omega_{i})
            \label{eq:QLcomponent}
\end{equation}
for atoms in a coordination shells, a coordination shell being a single peak of $g(r)$ (region where $g(r) > 1$), with $\Omega_{i}$ the vector from the center to atom $i$.  To determine the constant-volume melting point $T_{m}$ we heated crystals from $T=0.00$. At $T_{m}$, $g(r)$ crosses over from solid to liquid behavior and atoms become mobile.  The highest-melting-point crystal was body-centered-cubic, stable to $T_{m} \approx 1.08$. 

\begin{figure*}

\includegraphics{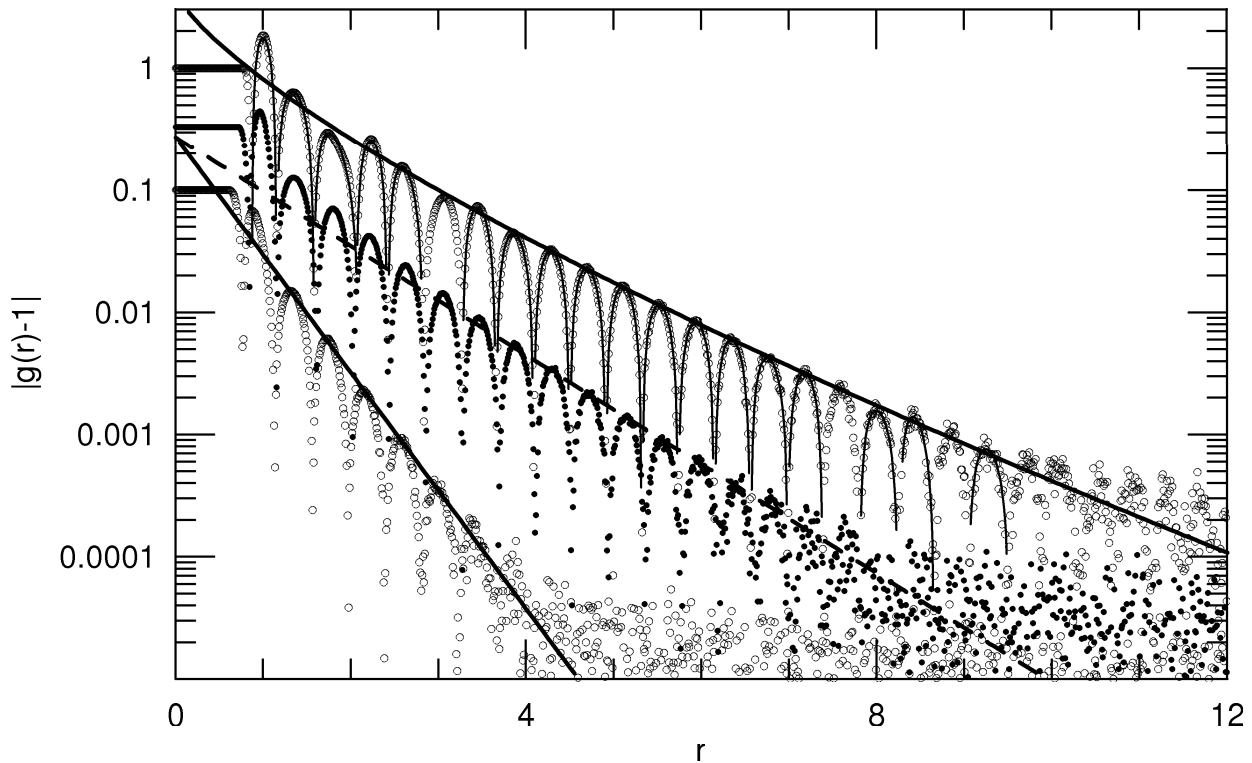}

\caption{\label{figure1} The all-atoms radial distribution function, plotted as $\mid g(r) -1\mid$ against $r$, at temperatures 20($\circ$), 2.01($\bullet$), and 0.56($\circ$).  $T=20$ and $T=2$ data are displaced downward for clarity; $\mid g(0)-1\mid = 1$ always.  Near-parabolas through $T=0.56$ data are exemplary fits of eq.\ \ref{eq:grform}, omitted for the ninth crest and third and tenth troughs to permit judgement of data quality.  Solid and dashed curves are envelopes based on eq.\ \ref{eq:stretched} and a simple exponential, respectively. The strong dependence of the range of liquid order on temperature is immediately transparent.}

\end{figure*}

Figure \ref{figure1} shows $\mid g(r)-1\mid$ at $T= 20$,  2.01, and 0.56.  Because we plot the magnitude, crests and troughs of $g(r)$ all appear as peaks. At the high temperature T=20, order persists through four crests and matching troughs, with no order for $r > 4$.  

  At $T=2$, at least eight crests (and intervening troughs) appear in $g(r)$.  At $T=0.56$ a dozen coordination shells are visible; ordering extends to $r \approx 11$.  This ordering is not microcrystallization: First, $T_{m}$ is 1.08, but ordering increases progressively starting at $T>10$. Second, features of $g(r)$ of warm crystals and the fluid do not match;
for example, the second crest of $g(r)$ is skewed toward small $r$ in crystals, and toward large $r$ in the fluid.  Simulations on small (say, $L \leq 20$) systems might overlook extended ordering because periodic boundary conditions would overlap coordination shells on opposite sides of each atom and obscure their forms. Indeed, we dropped preliminary simulations with $N=15,000$ to ensure we saw bulk properties.

Most crests and troughs of $g(r)$ fit accurately (but note Figure \ref{figure5}) to
\begin{equation}
      g(r) = T_{1} - T_{2} (r - T_{3})^{2} \exp( - T_{4} r).
      \label{eq:grform}
\end{equation}
$T_{1}$ is the peak height, $T_{3}$ is the peak center, and $T_{2}$ and $T_{4}$ are shape parameters; figure \ref{figure1} shows exemplary fits.

To characterize $\mid g(r) - 1 \mid$ we fit peak maxima to envelope functions, as done by Glotzer, et al.\cite{glotzer} with four-point two-time dynamic correlation functions. Our envelopes were an exponential $T_{1}^{\rm fit} (r) = t_{o} \exp(- \kappa r)$ (where $\kappa$ is an inverse range) and 
\begin{equation}
       T_{1}^{\rm fit2} (r) =  a \exp( - b r^{c}),
       \label{eq:stretched}
\end{equation}
with $a$, $b$, and $c$ as fit parameters.  The range of $T_{1}^{\rm fit}$ is defined to be  $\kappa^{-1}$ or by volume integrals  
\begin{equation}
    \langle R^{2}\rangle = \frac{\int d{\bf r} r^{2} T_{1}^{\rm fit2}(r) }{\int d{\bf r} T_{1}^{\rm fit2}(r)}.
    \label{eq:r2def}
\end{equation}  At large $T$, all peaks were fit.  For $T < 2$, fits were restricted to peaks with $r > 4$.  Crests and troughs of $g(r)$ all fit to one envelope.  

\begin{figure}

\includegraphics{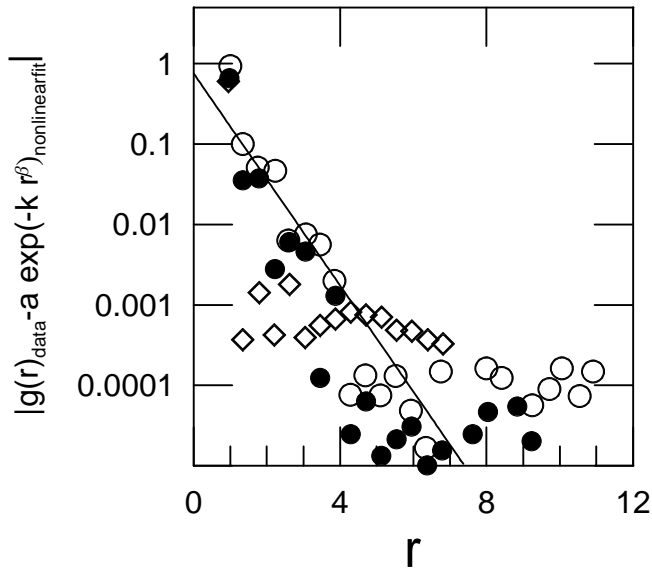}

\caption{\label{figure2} Deviations $\mid g(r) - T_{1}^{\rm fit2} (r) \mid$ of peaks of $g(r)$ at  temperatures 0.56 ($\circ$), 1.0 ($\bullet$), and 2.4, and (solid line) $\exp (- \kappa_{2} r)$ fit to $r \leq 4.0$ deviations at $T=0.56$. }

\end{figure}

The short-range ($r<4$) part of $g(r)$ of cold systems contains additional structure.  Figure \ref{figure2} shows deviation plots (peak height, minus envelope fitted for $r>4$) at three temperatures. For $r>4$ deviations are small. At $T=2.4$ there are no deviations.  Below $T=2$, deviations increase markedly toward small $r$. At all $T$, deviations stop at the fourth coordination shell and do not continue to larger $R$.  As a simple approximation we fit deviations to $\sim \exp(- \kappa_{2} r)$. The $\kappa_{2}^{-1}$ is a second length scale.  (To reject artifacts: For $T < 2$ we also fit eq.\ \ref{eq:stretched} to all peaks, finding relative to the $r>4$ fits a ten-fold larger error; the $r<4$ deviations remained.)  

\begin{figure}

\includegraphics{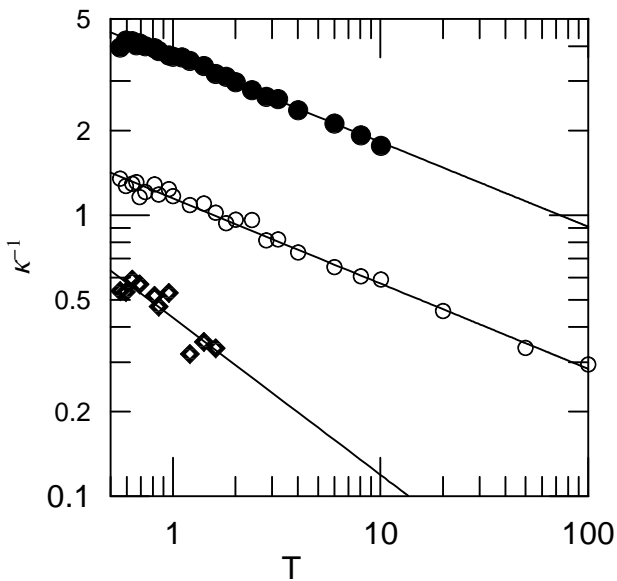}

\caption{\label{figure3} Temperature dependence of (top to bottom) $R$, $\kappa^{-1}$, and $\kappa_{2}^{-1}$, and fits to $\ell \sim T^{-x}$.}

\end{figure}

Figure \ref{figure3} shows $\kappa^{-1}$, $R$, and $\kappa_{2}^{-1}$ against $T$. $\kappa^{-1}$ and $R$ describe the outer coordination shells; they differ numerically because they weight large-$r$ effects differently. $\kappa_{2}^{-1}$ describes inner coordination shells of cold systems.  The increase in $\kappa_{2}^{-1}$ corresponds to $T$-dependences of inner peak heights, all within the temperature-independent maximum cluster radius.  These lengths scale as power laws 
\begin{equation} 
       \ell \sim T^{-x}.
       \label{eq:rangescale}
\end{equation}
for $x = 0.30, 0.30,$ and $0.56$, respectively. $\kappa^{-1}$ and $R$ share an exponent; they reflect the same length scale.  Exponents for $\kappa_{2}^{-1}$  and $R$ differ; $R$ and $\kappa_{2}$ represent physically distinct lengths. We have thus found two distinct static lengths in cold Lennard-Jones fluids, both seemingly diverging at low temperature, contrary to any belief that static lengths are quiescent at small $T$.    

\begin{figure}

\includegraphics{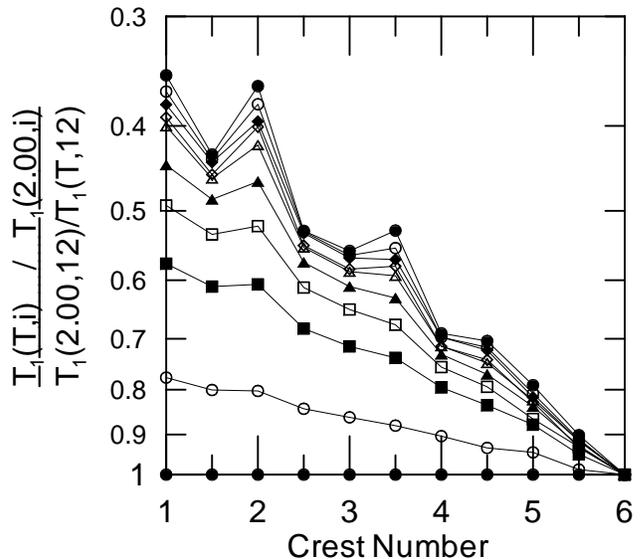}

\caption{\label{figure4} Temperature dependence of the normalized peak amplitude $T_{1}(T, i)$ ($i$ labelling the peaks of Figure \ref{figure1}, $i=1$ being the excluded-volume core, and $i=12$ being the sixth crest of $g(r)$, for the early crests (labelled on abscissa) and intervening troughs (unlabelled on abscissa) of $g(r)$, at temperatures (bottom to top) 2, 1.6, 1.2, 1.0, 0.86, 0.74, 0.69, 0.67, 0.65, and 0.55.  Note inverted vertical scale.}
\end{figure}

\begin{figure}

\includegraphics{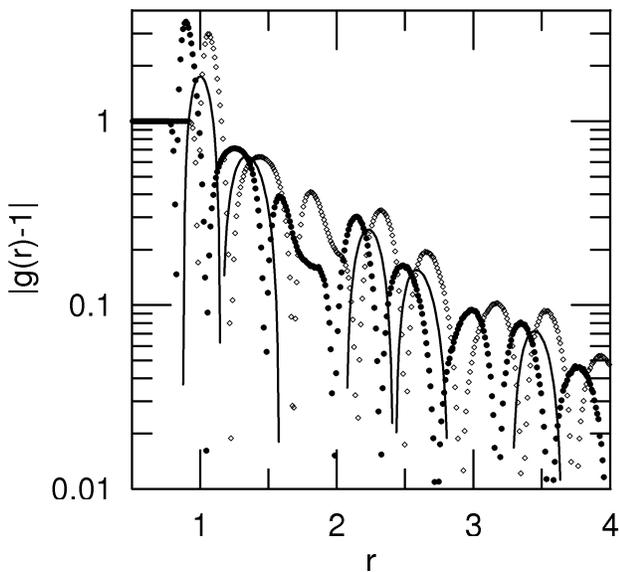}

\caption{\label{figure5} Expanded view of $g(r)$ (lines), $g_{AA}$ (open points), and $g_{BB}$ filled points) at $T=0.56$, showing emergent line structure.  $g(r)$ omitted for the second peak and third trough. }

\end{figure}

The scatter in Figure \ref{figure2} at $r < 4$ is non-random.  Figure \ref{figure4} shows amplitudes $T_{1}$  for the first peaks at various $T$.  Here $T_{1}$ is normalized: First, division by $T_{1}$ at $T=2$ suppresses peak to peak height variations.  Second, the $T$-dependence of the outer shells is suppressed by dividing out the $T$-dependence of the sixth crest of $g(r)$, which is the most accurately known of the outer ($r>5$) coordination shells.  Below $T=2$, the \emph{normalized} $T_{1}$ fall with decreasing $T$.  Below $T=1.0$, the first and second troughs and the second and third crests reach limiting amplitudes, but the first and second crests and third and fourth troughs continue to decrease.  It has long been known that the second peak has a characteristic glass-denoting characteristic, namely a notch and larger-$r$ shoulder.  We find weaker glass-denoting characteristics in other crests and troughs of $g(r)$.

Figure \ref{figure5} shows the detailed shapes of the inner crests and troughs of $g_{AA}(r)$ and $g_{BB}(r)$ at our coldest temperature, 0.56.  For both, the second crest (near $r=2$) has a strong larger-$r$ shoulder.  Estimating the skewness of the peaks by the half-widths at half-height, the third trough is markedly skewed to smaller $r$, and the first and third crests are skewed to larger $r$.  In contrast, at $T=2.01$ (not shown), the first crest and trough are slightly skewed, but the outer crests and troughs are not skewed appreciably.

\begin{figure}

\includegraphics{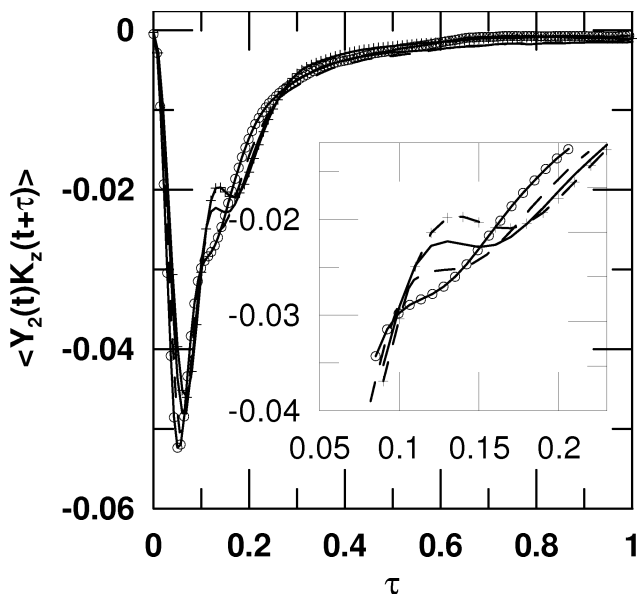}

\caption{\label{figure6} Dynamic correlation $\langle Y_{2}K\rangle$ between a spherical harmonic component of the local density of the first coordinate shell and the kinetic energy at temperatures 0.59($+$), 0.80 (solid line), 1.20(dashed line) and 2.00($\circ$).}

\end{figure}

We interpret the behavior of $g(r)$ as reflecting two independent effects.  First, there is a fundamental liquidlike order in $g(r)$ that packs atoms in regular shells.  As $T$ declines, the regularity of the shells increases, leading to shells becoming visible at larger $r$. The range $R$ of liquidlike ordering appears to increase indefinitely as $T$ is reduced, reaching $r \approx 11$ at $T=0.56$. 

Second, below $T=2$, short-range ordering appears.  We interpret this effect as arising from the formation of local clusters.  Clusters have a temperature-independent maximum extent, the fourth trough of $g(r)$.  Clusters are partially visualized in Figure \ref{figure4} ($T$ dependences of $T_{1}$ of crests one and two and trough three), and Figure \ref{figure5} (shoulders and asymmetries in the first three crests and third trough of $g(r)$ and the $g_{ii}(r)$).  

Third, clusters are transparent to long-wavelength scattering.  At $T=2$ and $T=0.56$, $\int_{0}^{4} (g(r)-1) d{\bf r} \approx 0$.  Cluster formation rearranges atomic positions for $r < 4$ but do not change the cluster mass.  

Within a sphere of diameter 8, a cluster might have 300 atoms.   However, all atoms in a sphere need not be part of a cluster.  Donati, et al.\cite{glotzer1} concluded from a four-point, two-time distribution function that the radial distribution function for pairs of immobile atoms, in a system and temperature close to ours, is relatively well-ordered at small distances and fades to zero beyond $r \approx 4$, the size of our clusters. Our static clusters are plausibly the immobile dynamic clusters of Donati, et al.

Observed clusters have properties predicted by Kivelson, et al.\cite{kivelson} for frustration-limited cluster-forming glasses, namely: Clusters (i) are stable at temperatures above the melting temperature, (ii) manifest frustration by having a temperature-independent maximum size, and (iii) have a local order orthogonal to the density.  Application of the Kivelson glass model is complicated by the low-$T$ divergence of the liquidlike length scale, and by $T$-dependent cluster internal structure seen in  $\kappa_{2}$.  

Finally, Kivelson, et al. predict that glass formation creates dynamic correlations between local order parameters orthogonal to the density, and time derivatives $\dot{A}$ of principle mechanical variables.  We evaluated $\langle Y_{2}K\rangle = \langle Q_{20}(0) K(\tau)\rangle$, where $Q_{20}(0)$ was computed for the first coordination shell, and $K(\tau) = p^{2}/2m$ is the kinetic energy of the central atom at later time $\tau$.  Figure \ref{figure6} shows that at small $T$ $\langle Y_{2}K\rangle$ gains (inset) new $\tau \neq 0$ structure, as predicted by Kivelson, et al.'s model\cite{kivelson}.

\begin{acknowledgments}

The partial support of this work by the National Science
Foundation under Grant DMR99-85782 is gratefully acknowledged.

\end{acknowledgments}

\end{document}